# Towards reliable use of artificial intelligence to classify otitis media using otoscopic images: Addressing bias and improving data quality

*Yixi Xu[1] and *Al-Rahim Habib[2,3,4]
Graeme Crossland[5]
Hemi Patel[5]
Chris Perry[6]
Kris Bock[7]
Tony Lian[2,3]
William B. Weeks[1]
Rahul Dodhia[1]
Juan Lavista Ferres[1]
Narinder Pal Singh[2,3], ORCID 0000-0002-7719-1832

*Co-first authors

The code is publicly available at https://github.com/microsoft/otoscope-img.data-bias-eval-and-active-labeling/.

**Affiliations**:

1. AI for Good Lab, Microsoft, Redmond, Washington, USA
2. Sydney Medical School, Faculty of Medicine and Health, University of Sydney, Camperdown, New South Wales, Australia
3. Department of Otolaryngology, Head and Neck Surgery, Westmead Hospital, Sydney, New South Wales, Australia
4. Department of Otolaryngology – Head and Neck Surgery, Queensland Children's Hospital, South Brisbane, Queensland, Australia
5. Department of Otolaryngology – Head and Neck Surgery, Royal Darwin Hospital, Tiwi, Northern Territory, Australia
6. University of Queensland Medical School, Brisbane, Queensland, Australia
7. Azure FastTrack Engineering, Microsoft, Brisbane, Queensland, Australia

## Abstract

Ear disease contributes significantly to global hearing loss, with recurrent otitis media being a primary preventable cause in children, impacting development. Artificial intelligence (AI) offers promise for early diagnosis via otoscopic image analysis, but dataset biases and inconsistencies limit model generalizability and reliability. This retrospective study systematically evaluated three public otoscopic image datasets (Chile; Ohio, USA; Türkiye) using quantitative and qualitative methods. Two counterfactual experiments were performed: (1) obscuring clinically relevant features to assess model reliance on non-clinical artifacts, and (2) evaluating the impact of hue, saturation, and value on diagnostic outcomes. Quantitative analysis revealed significant biases in the Chile and Ohio, USA datasets. Counterfactual Experiment I found high internal performance (AUC > 0.90) but poor external generalization, because of dataset-specific artifacts. The Türkiye dataset had fewer biases, with AUC decreasing from 0.86 to 0.65 as masking increased, suggesting higher reliance on clinically meaningful features. Counterfactual Experiment II identified common artifacts in the Chile and Ohio, USA datasets. A logistic regression model trained on clinically irrelevant features from the Chile dataset achieved high internal (AUC = 0.89) and external (Ohio, USA: AUC = 0.87) performance. Qualitative analysis identified redundancy in all the datasets and stylistic biases in the Ohio, USA dataset that correlated with clinical outcomes. In summary, dataset biases significantly compromise reliability and generalizability of AI-based otoscopic diagnostic models. Addressing these biases through standardized imaging protocols, diverse dataset inclusion, and improved labeling methods is crucial for developing robust AI solutions, improving high-quality healthcare access, and enhancing diagnostic accuracy.

## Introduction

Ear disease represents a significant global public health challenge, affecting individuals across all age groups and socioeconomic background.[1] It is a leading cause of disability, contributing to communication barriers, social isolation, and reduced quality of life. Among children, recurrent acute otitis media and chronic otitis media are preventable causes of hearing loss that can profoundly impact speech and language development, academic achievement, and long-term social integration.[2] Long-term hearing loss can hinder their ability to participate in social activities and achieve educational milestones, potentially reducing quality of life and future employment opportunities.[3] It is estimated that nearly 60% of hearing loss in children is due to avoidable causes such as vaccine-preventable diseases, ear infections, birth-related causes and ototoxic medicines.[1] In underserved populations, where access to timely healthcare is limited, these conditions often go undiagnosed or untreated, exacerbating health disparities.[3]

Timely diagnosis and intervention are critical for mitigating the effects of ear disease, yet access to specialty ear disease and hearing health services poses a significant barrier.[4] Many low- and middle-income countries have limited access to audiologists and otolaryngologists, leaving general practitioners and community health workers to manage complex cases with limited diagnostic tools.[5–7] For instance, only 56% of countries in the African region have one or more otolaryngologists per 1 million people, while 67% of European countries have more than 50 otolaryngologists per million population.[1] This disparity is particularly acute in rural and remote areas, where delays in diagnosis can lead to irreversible complications and poor otologic and hearing outcomes.[8]

To address these challenges, artificial intelligence (AI) and deep learning have emerged as promising adjunct tools for ear disease diagnosis. These technologies have been explored to analyse otoscopic images, with the aim of helping healthcare workers detect abnormalities, reducing diagnostic variability, and supporting triage in busy clinics or resource-constrained settings.[9] For example, AI models have shown potential in identifying conditions such as acute and chronic otitis media, offering healthcare workers a powerful tool to enhance diagnostic accuracy and streamline workflows.[10–14] However, the clinical implementation of these models is hindered by limitations in the existing literature, including biases in training datasets and challenges with generalizability.[15]

AI models excel at recognizing patterns within training data, sometimes unintentionally linking certain confounding features to clinical outcomes. For instance, melanoma detection models might rely on shortcuts based on surgical skin markings.[16] Unfortunately, assessing data bias has remained challenging and typically requires both domain expertise and technical ability.[17] Existing AI models in healthcare image analysis often perform well in controlled settings but may underperform in real-world clinical environments. Many models rely on datasets collected from a single institution, leading to biases that compromise their performance across diverse populations and settings. These biases, such as models learning irrelevant features like lighting conditions or camera settings, can undermine the reliability of diagnostic decisions.[17,18] Addressing these gaps is essential to ensure that AI tools improve patient outcomes rather than introduce new risks.

The aim of this study was to investigate the potential of deep learning to enhance otoscopic workflows by addressing the limitations of existing datasets. The specific objectives were to identify and analyse biases in publicly available otoscopic image datasets, develop practical guidelines to mitigate these biases and improve data collection practices. By focusing on the clinical utility of AI tools and their integration into real-world settings, this study provides a foundation for developing reliable, equitable, and impactful solutions to support clinicians and improve patient care.

## Methods

### Dataset Overview

We used three publicly available otoscopic image datasets in this study (Table 1). The Chile dataset comprised 880 images collected from 180 patients aged 7 to 65 years.[19] The Ohio, USA dataset included 454 images from the Ohio State University and Nationwide Children's Hospital.[20] The Türkiye dataset consisted of 956 images from the Ozel Van Akdamar Hospital.[21] Institutional Review Board (IRB) approvals were obtained for all datasets and were publicly accessible.[19–21] Adjustments were made for consistency: tympanostomy tube images and the 'other' category in the Türkiye dataset were excluded, retaining only tympanosclerosis images. For internal testing, the Türkiye and Chile datasets used predefined splits, while 20% of the Ohio, USA dataset was sampled using stratified random sampling.

### Bias Identification: Quantitative Assessment

#### Experiment I: Eclipse Extent

To evaluate whether AI models relied on irrelevant visual artifacts, a dataset manipulation technique was used. As shown in Figure 1, portions of the otoscopic images were covered with an elliptical black mask to obscure the tympanic membrane. A metric termed "Eclipse Extent" is the ratio of the mask's dimensions to those of the image. It quantifies the degree of masking, ranging from 0 (original, unmasked image) to 1 (almost fully obscured). Four deep learning models—ResNet-50, DenseNet-161, ViT-B-16, and ViT-B-16-384—were trained on the masked dataset (referred to as the "Eclipsed Dataset") and evaluated on their ability to classify normal and abnormal images both internally (within the same dataset) and externally (using different datasets). This experiment aimed to detect reliance on confounding features, such as background or lighting, that could artificially inflate model performance.

Experiment II: Image Saturation
To investigate the impact of image acquisition settings, otoscopic images were converted into hue, saturation, and value (HSV) color space. Two logistic regression models were developed using these color features: one trained on comprehensive HSV metrics (mean and standard deviation) and the other using only saturation variability (standard deviation). These models underwent internal and external validation to determine if diagnostic performance was influenced by image color characteristics, which are highly dependent on filming conditions (e.g., lighting, camera settings).

Bias Identification: Qualitative Assessment

Near-Duplicate and Stylistic Bias Analysis
To detect redundancy and stylistic biases in the datasets, distance-based clustering methods were employed. Three open-source datasets were split into five stratified folds for 5-fold cross-validation. Five ViT-B-16-384 models were trained, and their image feature embeddings were averaged. Cosine distance thresholds were used to group near-duplicate images (images that are visually nearly identical) and stylistically similar images. The clustering threshold (α) was adjusted iteratively to identify redundancies and stylistic correlations with clinical outcomes. This analysis highlighted instances where dataset biases, such as over-representation of certain image styles or redundant images, could undermine the generalizability of AI models.

Deep Learning Model Training

To illustrate the susceptibility of deep learning models to data bias, four distinct architectures were chosen: ResNet-50, a convolutional neural network utilizing residual connections to enable the training of deeper networks; DenseNet-161, another convolutional architecture noted for its dense connectivity where each layer is directly connected to all preceding layers; ViT-B-16, which employs the Transformer architecture directly on sequences of image patches; and ViT-B-16-384, a vision transformer with an input resolution of 384x384.
Data augmentation was applied during training including random resized cropping, horizontal and vertical flipping, color jitter, and elastic transformations. All models were trained using stochastic gradient descent with a learning rate of 0.01 and a batch size of 32 for 100 epochs. The final model used for evaluation was the model checkpoint saved at the 100th epoch. Model training was performed using Python on a single Nvidia V100 GPU.

Statistical Analysis

Model performance was evaluated using AUC and 95% confidence interval (CI) by DeLong's test. One-sided p values $< 0.05$ were considered statistically significant when comparing model performance on different subsets. For logistic regression models, the Wald method was used to derive the confidence interval for odds ratios and to assess the statistical significance of individual predictors. All statistical analysis was performed using R.

## Results

## Quantitative Bias Identification

### Counterfactual Experiment I: Eclipse Extent

Models trained on images with clinically relevant features obscured by elliptical masks (Eclipse Extent=1.0) demonstrated high internal performance (AUC > 0.90) for the Chile and Ohio, USA datasets, despite minimal view of the tympanic membrane being available (Table 2). However, these models performed poorly on external datasets, indicating reliance on dataset-specific artifacts, such as ear canal skin or lighting conditions, rather than clinically meaningful features. By contrast, the Türkiye dataset exhibited less bias. Using ViT-B-16-384 as an example, as masking increased (Eclipse Extent from 0 to 0.9 to 1.0), the internal AUC decreased from 0.88 (95% confidence intervals 0.83,0.94) to 0.62 (0.53, 0.71) to 0.53 (0.44, 0.62), suggesting a stronger reliance on meaningful diagnostic information.

### Counterfactual Experiment II: Image Saturation

Models trained on hue, saturation, and value (HSV) color features achieved high internal AUCs (Chile: 0.91 (0.86, 0.96), Ohio, USA: 0.92 (0.86, 0.99)) but showed reduced generalizability in external testing (Table 3). Abnormal images in the Chile and Ohio, USA datasets exhibited higher saturation variability, likely influenced by lighting conditions or camera settings during image acquisition. A logistic regression model trained using the saturation standard deviation (std) value on the Chile dataset showed high internal performance (AUC = 0.89 (0.83, 0.95)) and generalized effectively to the Ohio, USA dataset (AUC = 0.87 (0.84, 0.91)). A logistic regression model trained using the saturation std value on the Ohio, USA dataset performed well both internally (AUC = 0.86 (0.77, 0.95)) and externally on the Chile dataset (AUC = 0.85 (0.83, 0.88)) (Table 3). The Türkiye dataset displayed weaker correlations between saturation and clinical outcomes, resulting in lower internal AUCs (0.52 (0.43, 0.61)) but better external performance, indicating reduced reliance on artifacts.

## Qualitative Bias Identification

### Near-Duplicate Images

The Chile dataset contained 145 sets of near-duplicate images, which accounted for 61% of the dataset; 52% of the testing data had a near-duplicate copy in the training set (Table 4). These redundant images likely contributed to inflated internal performance by allowing models to memorize rather than generalize from the data. Model performance on testing samples with near-duplicates in the training set is significantly (p-value < 0.01) higher than the rest of the testing data regardless of model architecture and eclipse values (Supplementary Table 3). On average, across four model architectures, the difference in AUC increased from 0.08 to 0.26 (Eclipse Extent = 0.9). This suggests that models exploiting dataset-specific artifacts tend to overestimate performance on testing samples that have near-duplicates in the training set.

### Stylistic Biases

In the Ohio, USA dataset, two dominant image style categories were identified. Style I included 117 images, all classified as normal cases, while Style II comprised 90 images, predominantly representing effusion cases. These stylistic differences, driven by variations in imaging techniques and settings, strongly influenced model predictions, emphasizing the role of image acquisition protocols in shaping AI performance (Supplementary Figure 2).

## Discussion

We sought to evaluate the biases inherent in publicly available otoscopic image datasets and their impact on the performance of deep learning models for diagnosing middle ear conditions. Two counterfactual experiments focused on identifying biases linked to non-clinical features such as image masking and saturation. Models trained on the Chile and Ohio, USA datasets achieved high internal performance (AUC > 0.90) but failed to generalize to external datasets, largely due to reliance on dataset-specific artifacts such as lighting and ear canal skin patterns. In contrast, the Türkiye dataset demonstrated less bias, with internal AUCs decreasing as masking increased, indicating greater reliance on clinically meaningful features. Qualitative analyses revealed redundant and stylistically biased images, particularly in the Chile and Ohio, USA datasets, where duplication and stylistic correlations with clinical outcomes were evident. These findings underscore the critical influence of dataset bias on AI model performance and reliability.

The Ohio, USA dataset demonstrated a notable bias in image framing, with abnormal cases frequently focusing on specific regions of interest, such as partial views of the eardrum, while normal cases typically encompassed the entire tympanic membrane. This discrepancy allows models to exploit non-clinical cues, such as image framing, rather than clinically relevant features. To mitigate this form of bias, otoscopic images should consistently capture the entire tympanic membrane regardless of the diagnosis. The use of cameras equipped with automatic tympanic membrane detection systems could help standardize image quality while minimizing the need for extensive training for healthcare workers.

The second counterfactual experiment revealed that image saturation, a factor demonstrably influenced by lighting and camera settings, could effectively differentiate between normal and abnormal cases in both the Chile and Ohio, USA datasets (AUC > 0.85). This highlights the potential for bias introduced during image acquisition. This finding raises concerns that the gold standard of external validation may no longer be reliable if data bias is not adequately addressed. Ensuring that otoscope models and imaging protocols are independent of diagnostic outcomes and expert input is essential to reduce reliance on artifacts and enhance the clinical reliability of AI tools.

Analysis of the Chile dataset showed that a significant number of test images had near-identical counterparts in the training set, leading to artificially inflated performance metrics. For instance, models trained on masked (eclipsed) images still achieved high internal AUCs due to these redundancies, even when clinically relevant information was obscured. To address this, patient-based rather than image-based partitioning should be used when creating training and testing datasets to ensure a robust evaluation of model generalizability. Further, as a general quality check of publicly available datasets, duplicate images routinely should be eliminated prior to beginning analytic processes or splitting into training and testing datasets.

These findings have significant implications for the analysis of publicly available datasets, and, thereby, use of AI models in otoscopic diagnosis. For AI to become a reliable adjunct in clinical workflows, datasets must be curated to minimize biases, include diverse populations, and reflect varied imaging conditions. Rigorous external validation across geographically and demographically diverse datasets is essential to bridge the gap between research settings and real-world applications. This study highlights the need for standardized imaging protocols and improved data collection practices to enhance the reliability of AI models in otoscopy and elsewhere.

This study has several limitations. First, the focus on publicly available datasets may not encompass the full range of conditions encountered in routine clinical practice, including variation in otoscope devices, imaging protocols, and patient demographics. Second, the study's retrospective nature limits the ability to control for confounding factors related to image acquisition, including differences in operator technique, lighting conditions, and anatomical variations between adult and paediatric patients. Third, the methodology used to assess bias, specifically the application of an elliptical mask cantered on the tympanic membrane, does not fully reflect real-world diagnostic challenges. In clinical practice, abnormalities may be evaluated in specific anatomical regions of the tympanic membrane,[22] which were not specifically targeted in the masking approach. Future studies could explore region-specific masking techniques to better understand bias in AI model predictions. Fourth, the simulated masking and image saturation manipulations used in this study provide an artificial framework to evaluate bias but may not fully replicate the complexities of clinical image variability, such as motion artifacts, cerumen obstruction, or variability in clinician positioning. Finally, the reliance on a limited number of public datasets restricts the generalizability of findings, as publicly available datasets may not accurately reflect the diversity of patients, clinical environments, and healthcare settings globally. Expanding the analysis to include larger, more diverse datasets from multiple institutions and geographic regions would strengthen the applicability of these findings.

The strengths of this study lie in its comprehensive evaluation of dataset biases through both quantitative and qualitative methods. Counterfactual experiments combined with clustering-based analyses provided a nuanced understanding of factors influencing AI model performance. The inclusion of diverse datasets and systematic evaluation of non-clinical features, such as lighting and image style, offers valuable insights into the challenges of AI generalizability. The proposed data collection guidelines offer actionable recommendations for improving dataset quality and clinical applicability. Further, the tests conducted in this study may be considered as part of a pre-analytic data quality assessment to identify bias prior to model creation.

Future research should focus on mitigating dataset biases during model training through advanced techniques such as feature disentanglement and augmentation strategies that reduce reliance on confounding features.[23] Expanding the analysis to include larger and more diverse private datasets would provide a broader understanding of biases in otoscopic imaging. Incorporating multi-expert annotations and assessing inter-observer variability could further enhance dataset quality and reliability. Testing AI tools in real-world clinical environments with standardized imaging protocols will be critical for translating these findings into impactful healthcare solutions.

## Conclusion

This study revealed significant biases in publicly available otoscopic image datasets that impact the generalizability and reliability of AI models in diagnosing middle ear conditions. The findings demonstrate that AI models often rely on dataset-specific artifacts, such as lighting conditions and framing inconsistencies, rather than only clinical features in the region of the tympanic membrane, leading to compromised performance in external settings. Addressing these biases requires rigorous data curation, including the elimination of redundant images, standardized imaging protocols, patient-centered data partitioning to improve model generalizability, and rigorous quality assessment prior to embarking on analysis. Furthermore, ensuring consistency in image acquisition settings and implementing automated tympanic membrane detection can enhance diagnostic accuracy while reducing variability introduced by different operators and imaging equipment. Future research should focus on refining AI training strategies through feature disentanglement techniques, expanding dataset diversity to include varied populations and clinical settings, and integrating multi-expert annotations to improve data quality and reliability. These efforts are crucial for developing AI-based diagnostic tools that are robust, promote access to high-quality healthcare, and have clinical applicability across diverse healthcare environments.

**Acknowledgements:**

This research was supported by the Microsoft AI for Humanitarian Action Grant, the Ramsay Health Research Foundation Translational Challenge Grant, the Research Scholarship from the Garnett Passe and the Rodney Williams Memorial Foundation and Avant Foundation Doctor-in-Training Research Grant. We extend our gratitude to Anthony Ortiz and Zhongqi Miao for their insightful discussions that helped shape this research.

Figure 1: A sample otoscopic image and its eclipsed versions under varying Eclipse Extents of 0.0 (original), 0.9, and 1.0 from the Chile, Ohio, USA and Türkiye datasets.

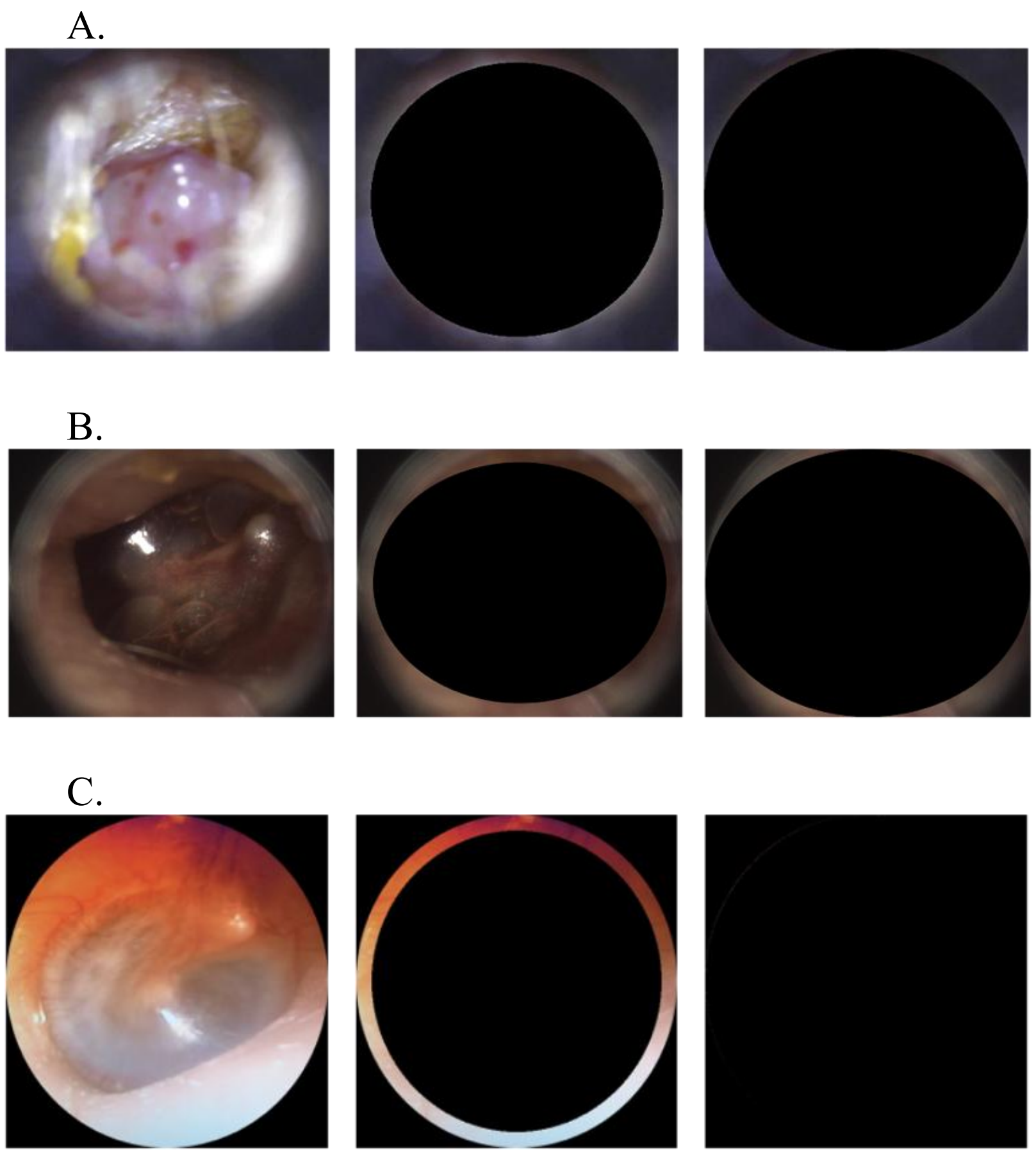

Legend: A – Chile dataset, B – Ohio, USA dataset, C – Türkiye dataset

Table 1: Image counts of sub-types in the otoscopy datasets.

| Dataset | Normal | Abnormal | | | | | |
|---|---|---|---|---|---|---|---|
| | | AOM | COM | Cerumen | Effusion | Myringosclerosis | Tympanosclerosis |
| Chile | 220 | 0 | 220 | 220 | 0 | 220 | 0 |
| Ohio, USA | 179 | 0 | 0 | 0 | 179 | 0 | 0 |
| Türkiye | 535 | 119 | 63 | 140 | 0 | 0 | 28 |
| Total | 934 | 119 | 283 | 360 | 179 | 220 | 28 |

Abbreviations: AOM (acute otitis media), COM (chronic otitis media)

Table 2. Internal and external performance of models trained on eclipsed images (Eclipse Extent = 0, 0.9, 1.0) from a single data source.

| Train Source | Eclipse Extent | Model Name | Internal Val AUC | External Testing AUC | | |
|---|---|---|---|---|---|---|
| | | | | Chile | Ohio, USA | Türkiye |
| Chile | 0.0 | densenet161 | 0.97 (0.93, 1) | N/A | 0.88 (0.85, 0.92) | 0.63 (0.6, 0.67) |
| | | resnet50 | 0.99 (0.96, 1) | N/A | 0.89 (0.86, 0.92) | 0.75 (0.71, 0.78) |
| | | vit_b_16 | 1 (0.99, 1) | N/A | 0.84 (0.8, 0.88) | 0.78 (0.75, 0.81) |
| | | vit_b_16_384 | 0.98 (0.96, 1) | N/A | 0.83 (0.79, 0.87) | 0.76 (0.72, 0.79) |
| | 0.9 | densenet161 | 0.94 (0.89, 0.99) | N/A | 0.66 (0.6, 0.72) | 0.45 (0.42, 0.49) |
| | | resnet50 | 0.96 (0.93, 0.99) | N/A | 0.68 (0.63, 0.74) | 0.59 (0.55, 0.63) |
| | | vit_b_16 | 0.94 (0.91, 0.98) | N/A | 0.58 (0.52, 0.64) | 0.41 (0.37, 0.45) |
| | | vit_b_16_384 | 0.97 (0.94, 0.99) | N/A | 0.39 (0.33, 0.45) | 0.43 (0.4, 0.47) |
| | 1.0 | densenet161 | 0.97 (0.94, 0.99) | N/A | 0.75 (0.7, 0.8) | 0.51 (0.47, 0.55) |
| | | resnet50 | 0.95 (0.92, 0.99) | N/A | 0.76 (0.7, 0.81) | 0.45 (0.41, 0.48) |
| | | vit_b_16 | 0.96 (0.92, 0.99) | N/A | 0.59 (0.53, 0.65) | 0.48 (0.44, 0.52) |
| | | vit_b_16_384 | 0.93 (0.89, 0.97) | N/A | 0.62 (0.56, 0.67) | 0.49 (0.45, 0.53) |
| Ohio, USA | 0.0 | densenet161 | 1 (0.99, 1) | 0.88 (0.86, 0.9) | N/A | 0.7 (0.66, 0.73) |
| | | resnet50 | 0.99 (0.98, 1) | 0.8 (0.76, 0.84) | N/A | 0.52 (0.48, 0.56) |
| | | vit_b_16 | 0.99 (0.96, 1) | 0.87 (0.85, 0.9) | N/A | 0.6 (0.56, 0.64) |
| | | vit_b_16_384 | 0.99 (0.97, 1) | 0.86 (0.83, 0.89) | N/A | 0.61 (0.57, 0.65) |
| | 0.9 | densenet161 | 0.98 (0.95, 1) | 0.35 (0.3, 0.39) | N/A | 0.49 (0.45, 0.53) |
| | | resnet50 | 0.96 (0.92, 1) | 0.4 (0.36, 0.44) | N/A | 0.48 (0.45, 0.52) |
| | | vit_b_16 | 0.94 (0.88, 1) | 0.57 (0.53, 0.61) | N/A | 0.52 (0.48, 0.56) |
| | | vit_b_16_384 | 0.93 (0.86, 1) | 0.56 (0.51, 0.6) | N/A | 0.4 (0.36, 0.44) |
| | 1.0 | densenet161 | 0.95 (0.9, 1) | 0.46 (0.42, 0.5) | N/A | 0.5 (0.46, 0.54) |
| | | resnet50 | 0.92 (0.85, 0.99) | 0.33 (0.29, 0.38) | N/A | 0.53 (0.49, 0.57) |
| | | vit_b_16 | 0.94 (0.87, 1) | 0.54 (0.5, 0.58) | N/A | 0.53 (0.49, 0.57) |
| | | vit_b_16_384 | 0.91 (0.85, 0.98) | 0.51 (0.47, 0.55) | N/A | 0.45 (0.41, 0.49) |
| Türkiye | 0.0 | densenet161 | 0.88 (0.82, 0.93) | 0.9 (0.87, 0.92) | 0.79 (0.74, 0.84) | N/A |
| | | resnet50 | 0.84 (0.77, 0.91) | 0.91 (0.9, 0.93) | 0.88 (0.84, 0.92) | N/A |
| | | vit_b_16 | 0.84 (0.78, 0.9) | 0.88 (0.86, 0.9) | 0.76 (0.71, 0.81) | N/A |
| | | vit_b_16_384 | 0.88 (0.83, 0.94) | 0.94 (0.92, 0.95) | 0.83 (0.79, 0.87) | N/A |
| | 0.9 | densenet161 | 0.58 (0.49, 0.67) | 0.44 (0.4, 0.48) | 0.69 (0.63, 0.74) | N/A |
| | | resnet50 | 0.72 (0.64, 0.8) | 0.8 (0.77, 0.83) | 0.62 (0.57, 0.68) | N/A |
| | | vit_b_16 | 0.68 (0.6, 0.76) | 0.61 (0.57, 0.65) | 0.44 (0.38, 0.5) | N/A |
| | | vit_b_16_384 | 0.62 (0.53, 0.71) | 0.66 (0.63, 0.7) | 0.65 (0.59, 0.7) | N/A |
| | 1.0 | densenet161 | 0.59 (0.5, 0.68) | 0.6 (0.56, 0.63) | 0.57 (0.5, 0.63) | N/A |
| | | resnet50 | 0.54 (0.45, 0.63) | 0.28 (0.24, 0.31) | 0.57 (0.51, 0.63) | N/A |
| | | vit_b_16 | 0.45 (0.36, 0.54) | 0.63 (0.59, 0.67) | 0.66 (0.6, 0.71) | N/A |
| | | vit_b_16_384 | 0.53 (0.44, 0.62) | 0.5 (0.45, 0.54) | 0.38 (0.32, 0.44) | N/A |

Table 3. Internal and external validation of logistic regression models using the HSV feature set and the single feature set.

| Train Source | Feature Set | Internal Val AUC | External Testing AUC | | |
|---|---|---|---|---|---|
| | | | Chile | Ohio, USA | Türkiye |
| Chile | HSV feature set | 0.91 (0.86, 0.96) | N/A | 0.76 (0.70, 0.81) | 0.66 (0.62, 0.69) |
| Ohio, USA | HSV feature set | 0.92 (0.86, 0.99) | 0.66 (0.62, 0.69) | N/A | 0.67 (0.64, 0.71) |
| Türkiye | HSV feature set | 0.58 (0.49, 0.67) | 0.42 (0.38, 0.46) | 0.81 (0.76, 0.85) | N/A |
| Chile | Single feature set | 0.89 (0.83, 0.95) | N/A | 0.87 (0.84, 0.91) | 0.66 (0.63, 0.70) |
| Ohio, USA | Single feature set | 0.86 (0.77, 0.95) | 0.85 (0.83, 0.88) | N/A | 0.66 (0.63, 0.70) |
| Türkiye | Single feature set | 0.52 (0.43, 0.61) | 0.85 (0.83, 0.88) | 0.87 (0.84, 0.91) | N/A |

Table 4: Statistics of identified near-duplicate image sets.

| Source | Set Count | Avg Size | Max Size | Redundant Count | Test Images with Near Duplicates in Training | | Test Images without Near Duplicates in Training | | Test Set Size |
|---|---|---|---|---|---|---|---|---|---|
| | | | | | Size | Abnormal ratio | Size | Abnormal ratio | |
| Chile | 145 | 5 | 17 | 536 | 84 | 0.69 | 76 | 0.81 | 160 |
| Ohio, USA | 8 | 2 | 3 | 10 | 6 | 0.33 | 66 | 0.52 | 72 |
| Türkiye | 92 | 2 | 3 | 94 | 18 | 0.67 | 154 | 0.34 | 172 |

**Supplements for**
Towards reliable use of artificial intelligence to classify otitis media using otoscopic images: Addressing bias and improving data quality

This document contains the following supplementary figures and tables:

Supplementary Figure 1: Near-duplicate image sets in the Chile, Ohio, USA and Türkiye datasets.

A. Chile dataset

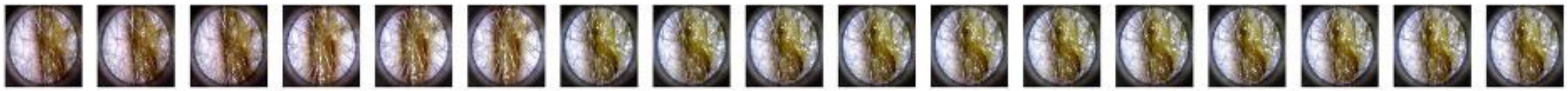

B. Ohio, USA dataset

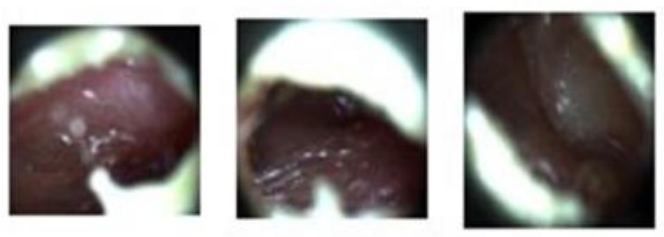

C. Türkiye dataset

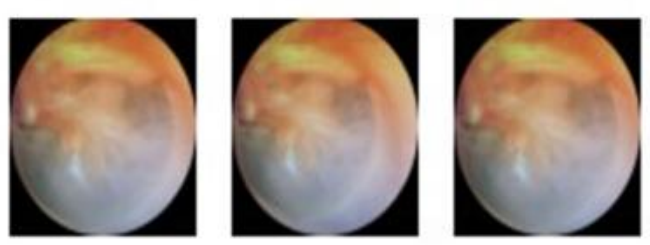

Supplementary Figure 2: One hundred random samples from the Style I set and the Style II set.

A. Style I

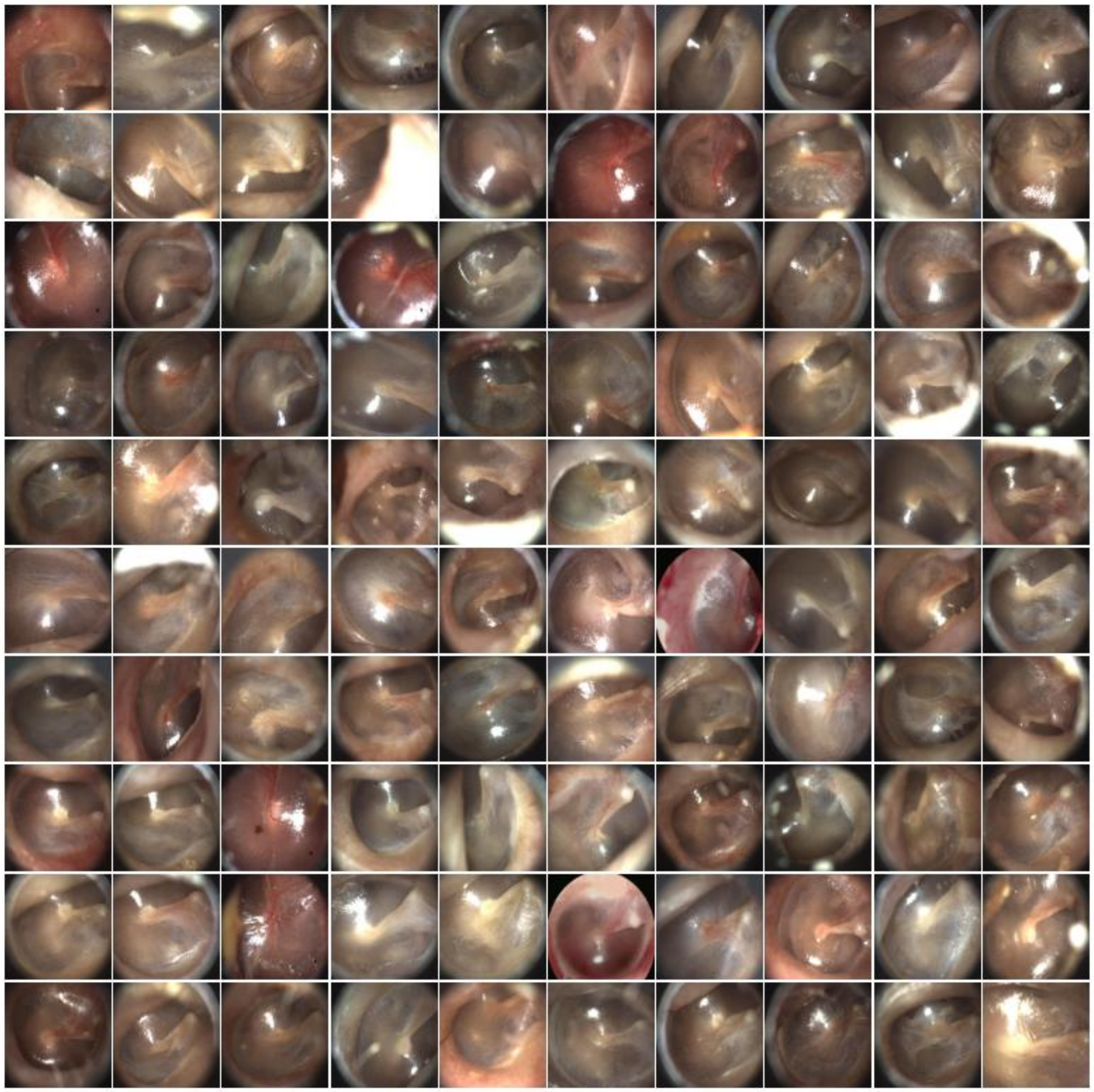

B. Style II

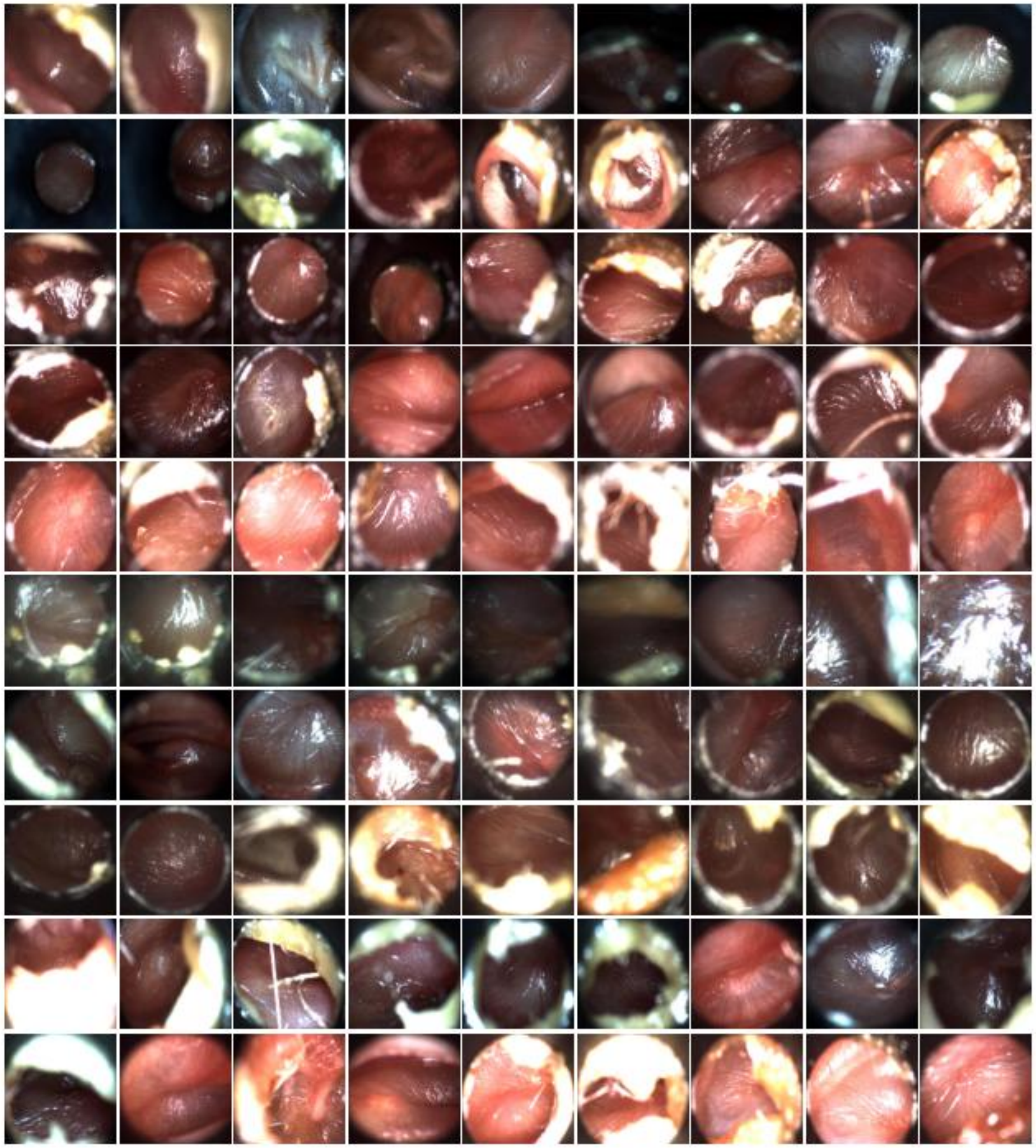

Supplementary Table 1: The odds ratio for saturation standard deviation in logistic regression models utilizing the single feature set.

| Train Source | Odds Ratio | P-Value |
|---|---|---|
| Chile | 1.22 (1.18, 1.27) | <0.01 |
| Ohio, USA | 1.25 (1.19,1.32) | <0.01 |
| Türkiye | 1.09 (1.07,1.12) | <0.01 |

Supplementary Table 2. Odds ratios for various variables in logistic regression models utilizing the HSV feature set.

| Train Source | Variable | Odds Ratio | P-Value |
|---|---|---|---|
| Chile | Hue mean | 0.99 (0.98, 1.00) | 0.14 |
| | Hue std | 0.99 (0.96, 1.01) | 0.25 |
| | Saturation mean | 1.02 (1.00, 1.04) | 0.04 |
| | Saturation std | 1.22 (1.16, 1.28) | <0.01 |
| | Value mean | 0.91 (0.89, 0.93) | <0.01 |
| | Value std | 1.09 (1.02, 1.16) | 0.01 |
| Ohio, USA | Hue mean | 1.04 (1.00, 1.09) | 0.03 |
| | Hue std | 1.02 (0.98, 1.05) | 0.31 |
| | Saturation mean | 1.02 (1.00, 1.04) | 0.02 |
| | Saturation std | 1.18 (1.11, 1.24) | <0.01 |
| | Value mean | 0.99 (0.97, 1.00) | 0.1 |
| | Value std | 1.05 (1.01, 1.09) | 0.02 |
| | Hue mean | 1.04 (1.01, 1.08) | 0.03 |
| Türkiye | Hue std | 0.98 (0.95, 1.02) | 0.3 |
| | Saturation mean | 1.05 (1.04, 1.07) | <0.01 |
| | Saturation std | 1.04 (1.01, 1.08) | 0.01 |
| | Value mean | 1.02 (0.99, 1.04) | 0.16 |
| | Value std | 1.02 (0.97, 1.07) | 0.5 |

Supplementary Table 3. Comparison of model performance on images with and without near duplicates in the Chile training set.

| Train Source | Eclipse Extent | Model Name | Internal Val AUC | Val Images with Near Duplicates in Training | Internal Val Images without Near Duplicates in Training | One-sided P-value |
|---|---|---|---|---|---|---|
| Chile | 0 | densenet161 | 0.97 (0.93, 1) | 0.99 (0.99, 1) | 0.88 (0.87, 0.9) | <0.01 |
| | | resnet50 | 0.99 (0.96, 1) | 0.99 (0.98, 1) | 0.9 (0.88, 0.91) | <0.01 |
| | | vit_b_16 | 1 (0.99, 1) | 0.99 (0.98, 1) | 0.93 (0.92, 0.94) | <0.01 |
| | | vit_b_16_384 | 0.98 (0.96, 1) | 0.99 (0.97, 1) | 0.92 (0.9, 0.93) | <0.01 |
| | 0.9 | densenet161 | 0.94 (0.89, 0.99) | 0.9 (0.83, 0.98) | 0.6 (0.58, 0.63) | <0.01 |
| | | resnet50 | 0.96 (0.93, 0.99) | 0.85 (0.77, 0.94) | 0.63 (0.6, 0.65) | <0.01 |
| | | vit_b_16 | 0.94 (0.91, 0.98) | 0.85 (0.75, 0.95) | 0.56 (0.53, 0.58) | <0.01 |
| | | vit_b_16_384 | 0.97 (0.94, 0.99) | 0.9 (0.83, 0.98) | 0.65 (0.62, 0.67) | <0.01 |
| | 1 | densenet161 | 0.97 (0.94, 0.99) | 0.93 (0.88, 0.99) | 0.82 (0.81, 0.84) | <0.01 |
| | | resnet50 | 0.95 (0.92, 0.99) | 0.92 (0.86, 0.98) | 0.73 (0.71, 0.75) | <0.01 |
| | | vit_b_16 | 0.96 (0.92, 0.99) | 0.91 (0.84, 0.99) | 0.78 (0.76, 0.8) | <0.01 |
| | | vit_b_16_384 | 0.93 (0.89, 0.97) | 0.91 (0.83, 0.98) | 0.77 (0.75, 0.79) | <0.01 |